\documentclass[]{spie}  
\pdfoutput=1

\newcommand\T{\rule{0pt}{2.6ex}} 
 
\usepackage{amsmath,amsfonts,amssymb}
\usepackage{graphicx}
\usepackage[colorlinks=true, allcolors=blue]{hyperref}
\usepackage{comment}

\title{LiteBIRD satellite: JAXA's new strategic L-class mission for all-sky surveys of cosmic microwave background polarization}

\author[1,2,3,4]{M.~Hazumi}
\author[5]{P.A.R.~Ade}
\author[6]{A.~Adler}
\author[7]{E.~Allys}
\author[8]{K.~Arnold}
\author[9]{D.~Auguste}
\author[10]{J.~Aumont}
\author[11]{R.~Aurlien}
\author[12]{J.~Austermann}
\author[13]{C.~Baccigalupi}
% ...
\affil[1]{High Energy Accelerator Research Organization (KEK), Tsukuba, Ibaraki 305-0801, Japan}
\affil[2]{Japan Aerospace Exploration Agency (JAXA), Institute of Space and Astronautical Science (ISAS), Sagamihara, Kanagawa 252-5210, Japan}
\affil[3]{Kavli Institute for the Physics and Mathematics of the Universe (Kavli IPMU, WPI), UTIAS, The University of Tokyo, Kashiwa, Chiba 277-8583, Japan}
\affil[4]{The Graduate University for Advanced Studies (SOKENDAI), Miura District, Kanagawa 240-0115, Hayama, Japan}
\affil[5]{Cardiff University, School of Physics and Astronomy, Cardiff CF10 3XQ, UK}
\affil[6]{Stockholm University}
\affil[7]{Laboratoire de Physique de l’$\acute{\rm E}$cole Normale Sup$\acute{\rm e}$rieure, ENS, Universit$\acute{\rm e}$ PSL, CNRS, Sorbonne Universit$\acute{\rm e}$, Universit$\acute{\rm e}$ de Paris, 75005 Paris, France}
\affil[8]{University of California, San Diego, Department of Physics, San Diego, CA 92093-0424, USA}
\affil[9]{Universit\'e Paris-Saclay, CNRS/IN2P3, IJCLab, 91405 Orsay, France}
\affil[10]{IRAP, Universit$\acute{\rm e}$ de Toulouse, CNRS, CNES, UPS, (Toulouse), France}
\affil[11]{University of Oslo, Institute of Theoretical Astrophysics, NO-0315 Oslo, Norway}
\affil[12]{National Institute of Standards and Technology (NIST), Boulder, Colorado 80305, USA}
\affil[13]{International School for Advanced Studies (SISSA), Via Bonomea 265, 34136, Trieste, Italy}
% ...

\authorinfo{Further author information: M.~Hazumi (E-mail: masashi.hazumi@kek.jp)}

\begin{document} 
\maketitle

\begin{abstract}
LiteBIRD, the Lite (Light) satellite for the study of $B$-mode polarization and Inflation from cosmic background Radiation Detection, is a space mission for primordial cosmology and fundamental physics.
JAXA selected LiteBIRD in May 2019 as a strategic large-class (L-class) mission, with its expected launch in the late 2020s using JAXA's H3 rocket.
LiteBIRD plans to map the cosmic microwave background (CMB) polarization over the full sky with unprecedented precision.
Its main scientific objective is to carry out a definitive search for the signal from cosmic inflation, either making a discovery or ruling out well-motivated inflationary models.
The measurements of LiteBIRD will also provide us with an insight into the quantum nature of gravity and other new physics beyond the standard models of particle physics and cosmology.
To this end, LiteBIRD will perform full-sky surveys for three years at the Sun-Earth Lagrangian point L2 for 15 frequency bands between 34 and 448\,GHz with three telescopes, to achieve a total sensitivity of 2.16\,$\mu$K-arcmin with a typical angular resolution of 0.5$^\circ$ at 100\,GHz. 
We provide an overview of the LiteBIRD project, including scientific objectives, mission requirements, top-level system requirements, operation concept, and expected scientific outcomes.
\end{abstract}


\section{Project Overview}
\label{sec:project}  
LiteBIRD, the Lite (Light) satellite for the study of $B$-mode polarization and Inflation from cosmic background Radiation Detection, is a space mission for primordial cosmology and fundamental physics.
After some initial conceptual studies\cite{Hazumi:2008zz,Hazumi:2011zz,Hazumi:2012gjy,Matsumura:2013aja,Matsumura:2014fte} that started in 2008,
we proposed LiteBIRD in 2015 as JAXA's large-class (L-class) mission candidate.
JAXA's L-class is for flagship science missions with a 300\,M USD cost cap.
There will be three L-class missions in about ten years launched using JAXA's H3 rocket.
LiteBIRD passed an initial down-selection and in 2018 completed a two-year Pre-Phase-A2 concept development phase.
JAXA selected LiteBIRD in May 2019 as the second L-class mission after
MMX, the Martian Moons Exploration, which will be launched around 2025.

The LiteBIRD Joint Study Group has more than 250 researchers from Japan, North America, and Europe with experience in CMB experiments, X-ray satellite missions, and other large projects in high-energy physics and astronomy.
In particular, a large number of researchers who worked on the Planck satellite are members of LiteBIRD.
We thus consider LiteBIRD to be the successor to the Planck satellite.

The main scientific objective of LiteBIRD is to carry out a definitive search for the signal from cosmic inflation, either making a discovery or ruling out well-motivated inflationary models. 
The measurements of LiteBIRD will also provide us with insight into the quantum nature of gravity and other new physics beyond the standard models of particle physics and cosmology. 
To this end, LiteBIRD plans to map the CMB polarization over the full sky with unprecedented precision. 

Although the hot Big-Bang picture is well supported by many distinct types of observation, several critical `origins' problems remain unanswered. The leading theory today to resolve these problems is cosmological inflation, hypothesizing that our Universe went through an accelerating expansion phase at very early stages, effectively beginning the hot Big Bang.
The cosmic inflation hypothesis predicts the emission of primordial gravitational waves during the inflationary era. 
These primordial gravitational waves should have imprinted a unique signature, called the $B$ modes, in the polarization pattern of the CMB\cite{Kamionkowski:1996zd, Seljak:1996gy, Zaldarriaga:1996xe, Kamionkowski:1996ks}.
Measurements of the large-angle CMB polarization are known as the most sensitive probe for primordial gravitational waves. 
State-of-the-art technology is required for detection, since the $B$-mode signal will be much fainter than the already-detected $E$-mode pattern.

The primordial $B$-mode measurements with LiteBIRD will also be the first stringent test of quantum gravity, which should exist behind any inflationary model.
Here `quantum gravity' means a theory that copes in a single framework with two pillars of physics: (1) Einstein's theory of general relativity that describes gravity; and (2) quantum mechanics. 

At this SPIE 2020 conference, there are several contributions with more details on the design of individual components of the LiteBIRD satellite\cite{Sekimoto2020,Montier2020,Westbrook2020,Sakurai2020,HTakakura2020,Tsuji2020,Tominaga2020,lamagna2021}.
The purpose of this article is to give a concise overview of LiteBIRD as an introduction to the other LiteBIRD proceedings.
In Sect.~\ref{sec:mission_requirements}, we describe our Level-1 mission requirements, or scientific requirements, and the rationale behind them.
In Sect.~\ref{sec:system_requirements}, we introduce our requirements flow and explain the measurement requirements.
After describing the launch vehicle (Sect.~\ref{sec:rocket}), the spacecraft (Sect.~\ref{sec:spacecraft}), the payload module (Sect.~\ref{sec:payload}), and the operation concept (Sect.~\ref{sec:operation}), we discuss the expected scientific outcomes in Sect.~\ref{sec:outcomes} and give a summary in Sect.~\ref{sec:summary}.

\section{Science Requirements}
\label{sec:mission_requirements}
\begin{figure}[htb!]
\centering
\includegraphics[width=1.0\textwidth]{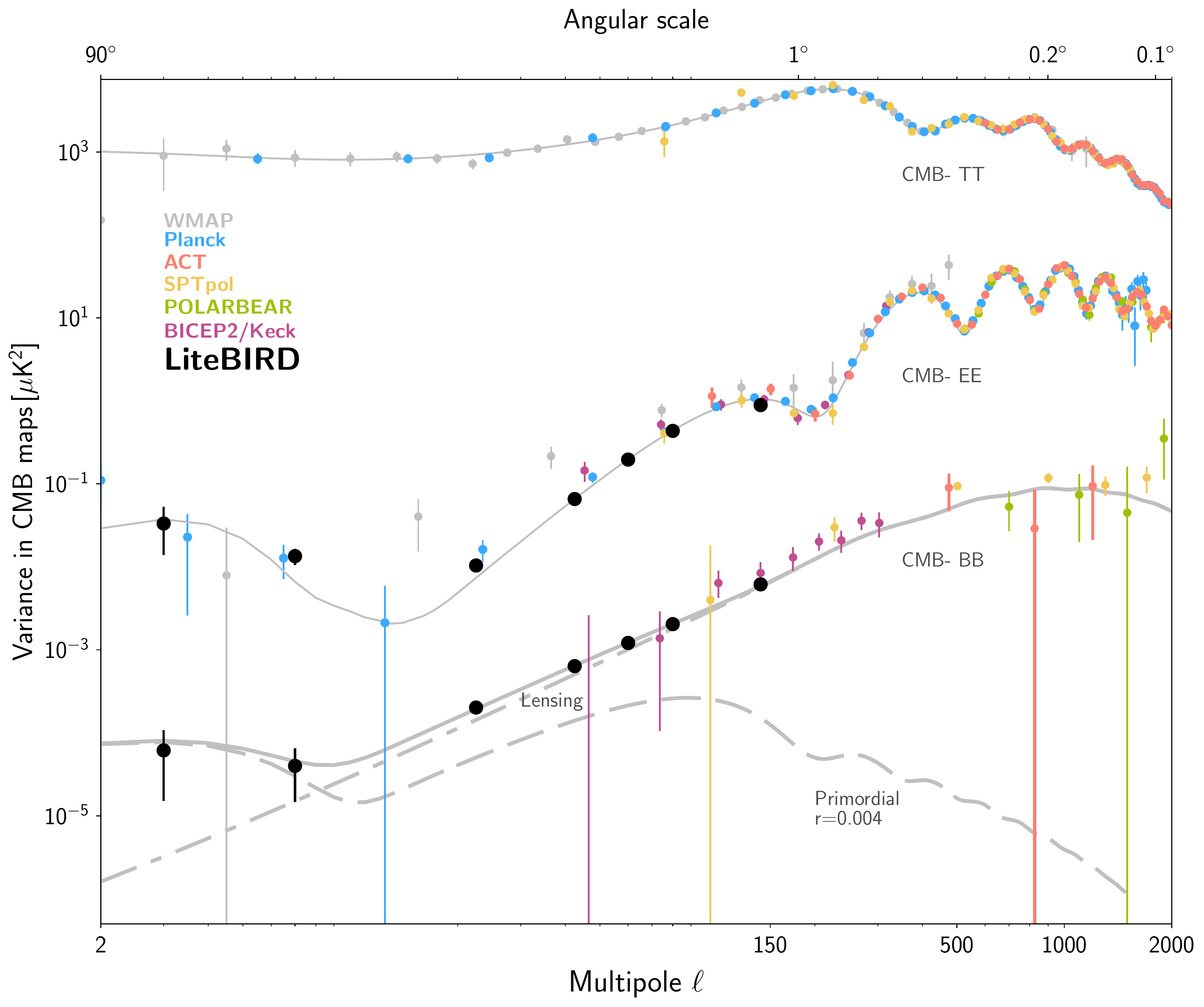}
\caption{Summary of present measurements of CMB power spectra\cite{Hinshaw:2012aka,Bennett:2012zja,Ade:2017uvt,Henning:2017nuy,Ade:2018gkx,Aghanim:2019ame,Sayre:2019dic,Aiola:2020azj,Choi:2020ccd,Adachi:2020knf} and expected polarization sensitivity of LiteBIRD.}
\label{fig:clbb}
\end{figure} 
In Fig.~\ref{fig:clbb} we summarize the present measurements of the CMB power spectra, including $B$ modes, with
the expected polarization sensitivities of LiteBIRD displayed.
The $B$-mode power is proportional to the tensor-to-scalar ratio, $r$, which is observationally constrained
to be $r < 0.06$ (95\%C.L.)\cite{Ade:2018gkx}, with a recent update using Planck data to $r < 0.044$\cite{Tristram2020}. 
The next-generation of CMB polarization experiments on the ground have the potential to see a hint of the signal around $\ell \sim 100$,
coming from the recombination epoch.
However, if $r$ is less than approximately 0.03, the $B$ modes due to gravitational lensing become dominant.
Removing contamination of the lensing $B$ modes, often called `delensing', is needed in this case.
In contrast, another excess at $\ell < 10$, which is due to reionization, is larger than
the lensing $B$ modes, even at $r=0.001$.
In order to access the reionization peak, one needs to survey the full sky, where
the advantage of observing in space is clear.

The critical question is: to what precision should $r$ be measured?
Here we introduce the total uncertainty on $r$, $\delta r$, 
which consists of five components: (instrumental) statistical uncertainties; systematic uncertainties; uncertainties due to contamination of foreground components; uncertainties due to gravitational lensing; and uncertainties due to observer biases.
There are many different inflationary models under active discussion, which predict different values of $r$.
Among them, there are well-motivated inflationary models that predict $r > 0.01$\cite{Kamionkowski:2015yta}.
If our requirement is $\delta r < 0.001$, we can provide more than $10\,\sigma$ detection significance for such models. On the other hand,
if LiteBIRD finds no primordial $B$ modes and obtains an upper limit on $r$,
this limit would be stringent enough to set severe constraints on the physics of inflation.
As discussed in Ref.~\citenum{Linde:2016hbb},
if we obtain an upper limit at $r\sim 0.003$, we can completely rule out
one important category of models, namely any single-field model in which
the characteristic field-variation scale of the inflaton potential is greater than the reduced Planck mass.

\begin{table}[htb!]
\centering
\begin{tabular}{|lp{9.5em}|p{33em}|} 
\hline
ID	& Title & Requirement description\\
\hline
Lv1.01 & Tensor-to-scalar ratio $r$ measurement sensitivity & 
The mission shall measure $r$ with a total uncertainty of $\delta r\,{<}\,1 \times 10^{-3}$.\T
This value shall include contributions from instrument statistical noise fluctuations, instrumental systematics, 
residual foregrounds, lensing $B$ modes, and observer bias, and shall not rely on future external data sets.\\
\hline 
Lv1.02 & Polarization angular power spectrum measurement capability &
The mission shall obtain full-sky CMB linear polarization maps for achieving ${>}\,5\sigma$ significance 
using $2\,{\leq}\,\ell\,{\leq}\,10$ and $11\,{\leq}\,\ell\,{\leq}\,200$ separately, assuming $r\,{=}\,0.01$. 
We adopt a fiducial optical depth of $\tau=0.05$ for this calculation.\\
\hline
\end{tabular}
\caption{Two science requirements of LiteBIRD, also called Level~1 (Lv1) mission requirements.}
\label{tbl:lv1}
\end{table}
Based on all the considerations described above,
we decided to impose the requirements described in Table~\ref{tbl:lv1}.
The first, Lv1.01, shall be achieved without delensing using external data;
if external data are available, we may further reduce $\delta r$\cite{Diego-Palazuelos2020}.
The second requirement, Lv1.02, becomes essential when $r$ is large.
If there is some indication of the primordial $B$ modes before observations by LiteBIRD, 
that would imply a relatively large value of $r$. In this case,
data from LiteBIRD will allow us to measure the $B$-mode signals from reionization and recombination
simultaneously.
If the spectral shape is consistent with the expectation
from the standard cosmology, that will narrow down the list of possible inflationary models, and
provide a much deeper insight into the correct model. 
If we observe an unexpected power spectrum beyond the standard model prediction,
that will lead to a revolution in our picture of the Universe.      
Lv1.02 also sets the angular resolution requirement for LiteBIRD.

\section{Measurement Requirements and System Requirements}
\label{sec:system_requirements}
To satisfy the science requirements described in the previous section, we use the requirements flow-down framework shown in Table~\ref{tab:reqflow}. 
To derive Lv2 measurement requirements from Lv1 science requirements,
we also consider program-level constraints, such as the cost cap, which are not controlled by the LiteBIRD team.
We also use agreed-upon assumptions between the LiteBIRD team and other parties or within the LiteBIRD team. Examples include assumptions on the complexity of the astronomical foreground components, the cooling-chain lifetime, and basic system redundancy guidelines.
There are in total 11 Lv2 measurement requirements on the statistical uncertainty (Lv2.01), the systematic uncertainty (Lv2.02), the scan strategy (Lv2.03), the angular resolution (Lv2.04), calibration measurements (Lv2.05), error budget allocation (Lv2.06), systematic error budget allocation (Lv2.07), the duration of the normal observation phase (Lv2.08), the orbit (Lv2.09), observer bias (Lv2.10), and noise-covariance knowledge (Lv2.11).
Our error budget (Lv2.06) is defined such that an equal amount, $(1/\sqrt{3})\times 10^{-3} = 0.57\times 10^{-3}$, is given to each of the following three components: the total statistical error after foreground separation $\sigma_{\rm stat}$; the total systematic error $\sigma_{\rm syst}$; and a margin. 
The requirements are thus $\sigma_{\rm stat} < 0.57\times 10^{-3}$ on the statistical uncertainty (Lv2.01) and $\sigma_{\rm syst} < 0.57\times 10^{-3}$ on the systematic uncertainty (Lv2.02).
Since we assume no delensing using external data, $\sigma_{\rm stat}$ includes uncertainties from the lensing $B$-mode component. 
Uncertainties due to foreground separation are also in $\sigma_{\rm stat}$.
The observer bias (Lv2.10) shall be much smaller than $\sigma_{\rm syst}$.
The requirement on the statistical uncertainty (Lv2.01) has six sub-requirements on
(1) the measurement on CMB sensitivity, (2) on dust emission, and (3) on synchrotron emission, 
(4) separation of CO lines, (5) the number of observing bands, and (6) the observing frequency range.
These are determined through  detailed simulation (described in Sect.~\ref{sec:outcomes}).
We require full-sky surveys (Lv2.03) to obtain the $B$ modes to the lowest multipole of $\ell = 2$.
The angular resolution (Lv2.04) shall be better than 80\,arcmin (FWHM) at the lowest frequency band
in order to perform precision measurements at $\ell\,{=}\,200$.
The regular observation phase (Lv2.08) shall be three years, considering the total cost cap and cooling-chain lifetime.
The orbit (Lv2.09) shall be a Lissajous orbit around the Sun-Earth L2 point to avoid the influence of the Sun, Moon, or Earth radiation (discussed further in Sect.~\ref{sec:operation}).
Requirements on calibration measurements (Lv2.05, Lv2.11) and systematic error budget allocation (Lv2.07) will be explained in Sects.~\ref{sec:payload} and \ref{sec:outcomes}, respectively.

Lv1 and Lv2 requirements are collectively called `mission requirements.'
In general, several possible designs meet mission requirements.
We, therefore, performed implementation trade-off studies to choose the best design.
Here, we also consider the program-level constraints and assumptions that we used to set Lv2 requirements.

Lv3 instrument requirements constitute top-level system requirements.
An essential distinction between Lv2 and Lv3 is that Lv3 instrument requirements are for the instrument chosen from trade-off studies, while Lv2 measurement requirements do not assume a specific instrument in principle. 
Lv3 requirements include general system requirements not only for mission instruments but also for the bus system,\footnote{Also called the `service module,' or `SVM' for short.} ground segments and ground-support equipment.
There are too many Lv3 requirements to list here.
The requirement flow's tree structure is also too detailed to show, since some Lv3 requirements derive from more than one Lv2 requirement; however, we will explain some essential Lv3 requirements in Sect.~\ref{sec:payload}.
%
\begin{table}[htb!]
\centering
	\begin{tabular}{|p{0.15\textwidth}|p{0.18\textwidth}|p{0.57\textwidth}|}
	\hline
	   \multicolumn{1}{|c|}{Class}
	& \multicolumn{1}{c|}{Symbol}
	& \multicolumn{1}{c|}{Description}\\ \hline\hline
	%
	\multicolumn{3}{|c|}{Mission requirements}\\
	\hline
	Level~1 (Lv1)     & Lv1.XX                 & Top-level quantitative science requirements that are\\ 
	science            & (e.g., Lv1.01)        & directly connected to the full success of the mission.\\
	requirements   &                            & \\ 
	\hline
	Level~2 (Lv2)     & Lv2.XX(.YY) (e.g. & Measurement requirements to achieve Lv1.\\
	measurement  & Lv2.01, Lv2.01.01) & No assumption is made on an instrument.\\
	requirements   &                            & \\ 	
	\hline
	%
	\multicolumn{3}{c}{$\downdownarrows$}\\
	\multicolumn{3}{c}{Implementation trade-off studies}\\
	\multicolumn{3}{c}{$\downdownarrows$}\\
	\hline
	\multicolumn{3}{|c|}{System requirements}\\
	\hline
	Level~3 (Lv3)     & Lv3.XX(.YY) (e.g. & Top-level implementation requirements for a chosen\\
	instrument       & Lv3.01, Lv3.01.01)  & instrument to achieve Lv2. Between Lv2 and Lv3 are\\
	requirements   &                             & tradeoff studies for instrument selection.\\
	\hline   
	Level~4 (Lv4)     & Lv4.XX(.YY) (e.g. & Component-level requirements to achieve Lv3.\\
	component      & Lv4.01, Lv4.01.01)  & \\
	requirements   &                             & \\
	\hline   
	Level~5 (Lv5)     & Lv5.XX(.YY) (e.g. & Sub-level build specifications to achieve Lv4.\\
	Sub-level build & Lv5.01, Lv5.01.01)  & \\
	specifications   &                             & \\
	\hline   
  	\end{tabular}
\caption{Definitions of five requirement levels used in LiteBIRD's requirements flow-down.
We split the requirements into five levels, from the top-level science requirements (Lv1) to sub-level build specifications (Lv5).
Each level is allowed to have a sub-structure; for example, a Level~2 requirement Lv2.01 has six sub-requirements (such as Lv2.01.01, Lv2.01.02).}
\label{tab:reqflow}
\end{table}

\section{Launch Vehicle}
\label{sec:rocket}
LiteBIRD will be launched on an H3\cite{JAXA_H3}, Japan's new flagship rocket.
It will achieve high flexibility, high reliability, and high performance at a lower cost than the currently used H-IIA rocket.
The H3 rocket is under development with its prime contractor, Mitsubishi Heavy Industries, with a maiden flight scheduled in the Japanese fiscal year of 2021.
The first stage of the H3 rocket will adopt the newly-developed liquid engine, LE-9, which achieves a 1.4 times larger thrust than the LE-7A engine currently in use.
Its second-stage engine, LE-5B-3, and the solid rocket booster, SRB-3, will also be improved.
The launch capability of the H3 rocket to the geostationary transfer orbit will be the highest ever among JAXA's launch vehicles,
exceeding that of the existing H-IIA and H-IIB launch vehicles.
The launch facility at Tanegashima Space Center will also be upgraded following the development of H3.

The design of the H3 rocket allows for several different configurations.
The rocket type is defined by the combination of the number of first-stage engines (2 or 3), the number of
solid rocket boosters (0, 2, or 4), and the length of the fairing (short or long).\footnote{Some of the combinations may not be offered as a standard lineup, based on market research.}
These lineups make it possible to cope with various payload sizes and orbits.
Considering the size, weight, and orbit of LiteBIRD, we plan to adopt the H3-22L configuration,
which means two first-stage engines, two boosters, and the long fairing.

The H3 rocket is designed to have launch capability of at least 4\,t to the Sun-synchronous orbit (500\,km in altitude),
and 6.5\,t to the geostationary transfer orbit.
The parameters of these orbits, however, are not appropriate for estimating the capability to L2.
We thus use the C3-based launch capability to evaluate the maximum-allowed weight for the L2 orbit.
Here, C3 is defined as a square of the residual velocity, which the payload launched from the Earth possesses at infinity.
The launch capability defined for ${\rm C3}\,{=}\,0$ is a good approximation for L2.
It is noteworthy that the launch capability vastly changes with the number of solid rocket boosters,
whereas the number of main engines has only a moderate impact on the launch capability.
Selection of the fairing also has little impact on the launch capability.
With no solid rocket booster, e.g., with H3-30S, the launch capability for ${\rm C3}\,{=}\,0$ is far less than 2\,t, 
which is much smaller than what is required for LiteBIRD.
On the other hand, if two solid rocket boosters are used, e.g., with H3-22L or H3-32L, 
the launch capability (${\rm C3}\,{=}\,0$) becomes larger than 3.5\,t, which is sufficient for LiteBIRD.
This launch capability changes slightly with the selection of the number of main engines.
Because two main engines can afford sufficient launch capability, 
we select H3-22L for the launch vehicle of LiteBIRD.
Note that the long fairing is necessary for LiteBIRD to fit.
Considering the estimated launch capability (${\rm C3}\,{=}\,0$) of H3-22L and the current estimation of the weight of LiteBIRD,
together with the various ambiguities of the estimations, we set the provisional requirement on the total
weight of LiteBIRD as ${<}\,3.5$\,t.
This requirement may be updated after the first flight of the H3 rocket.

In most cases, the launch environment of H3 is expected to be similar or more moderate than that of H-IIA.
Details of the launch environment may depend on the rocket configuration, especially on the number of solid rocket boosters,
the satellite mass, and the flight path.
We assume the launch environment of H-IIA in general for the design of LiteBIRD, to be on the safe side.
However, when the launch environment is critical in the design, such as for the mechanical requirement on the fundamental frequency
of the satellite, we adopt the requirements based on the current best estimation of the performance of H3.

\section{Spacecraft}
\label{sec:spacecraft}
\begin{figure}[htb!]
\centering
\includegraphics[width=1\textwidth]{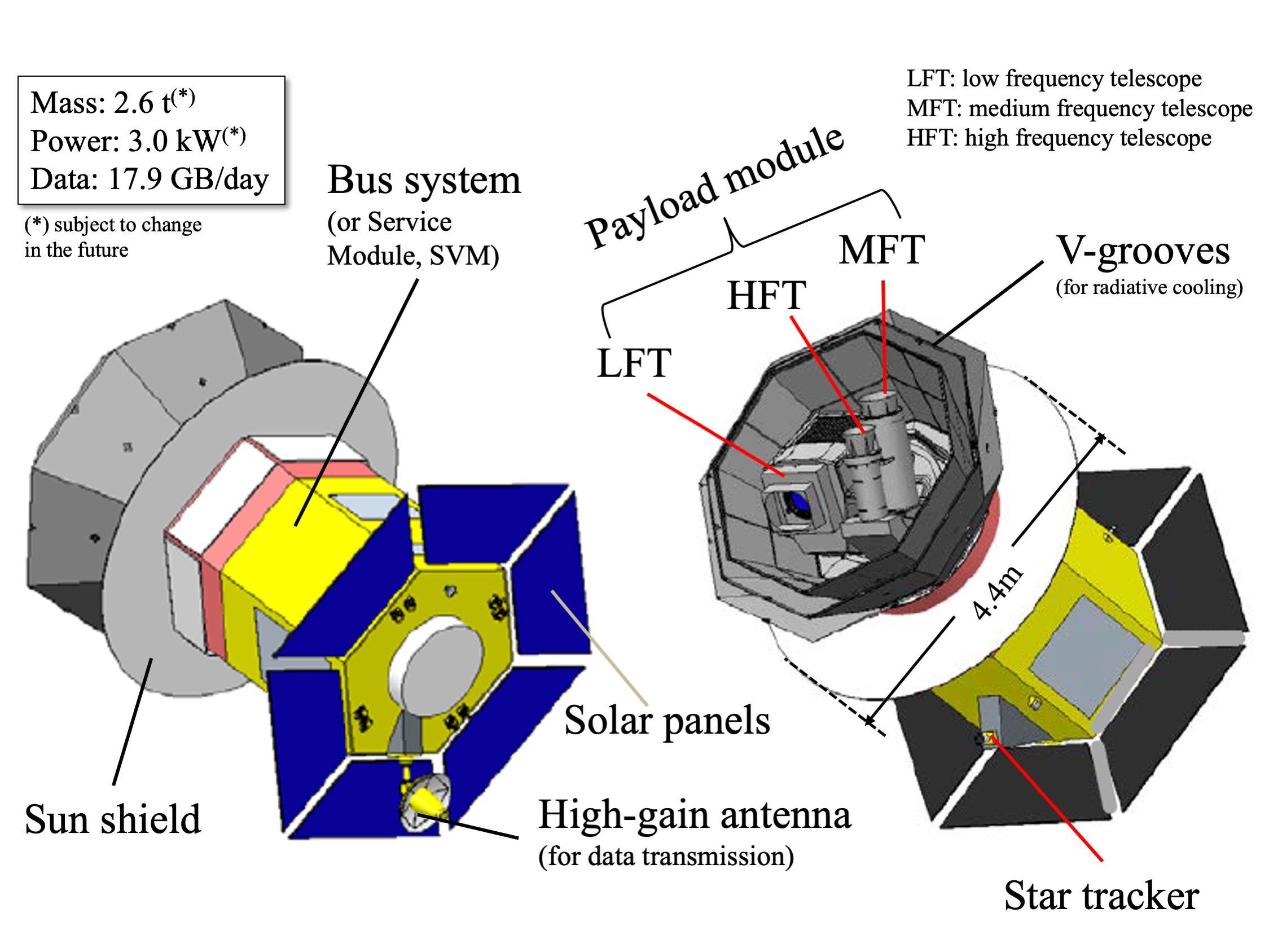}
\caption{Conceptual design of the LiteBIRD spacecraft. The payload module (PLM) houses the low-frequency telescope (LFT),
the medium-frequency telescope (MFT), and the high-frequency telescope (HFT).}
\label{fig:satellite_overview} 
\end{figure}

\noindent
The overall structure of the spacecraft for LiteBIRD is determined directly from the mission requirements.
The axisymmetric shape of the spacecraft is selected to make the spin easier.
Because the satellite's spin axis should be near the Earth-Sun line,
it is natural to place the telescopes 
and the solar panels at the opposite ends of the spacecraft.
We chose to place the PLM (including the telescopes) at the top of the spacecraft and the
solar panels at its bottom, perpendicular to the spin axis. 
The high-gain antenna should be placed on the bottom side of the satellite, i.e., opposite the mission instruments, to point to the Earth and reduce interference with the telescopes. 
Based on these considerations, we show the basic structure of the spacecraft in Fig.~\ref{fig:satellite_overview}.

In this configuration, the whole spacecraft spins, and the possibility of using a slip-ring to rotate only the PLM is {\it not\/} adopted.
The main reasons for this selection are to handle large heat dissipation in the PLM and to reduce the possibility of a single-point failure.
The PLM is equipped with mechanical coolers, which dissipate a fairly large amount of heat.
Sufficient radiator size to dissipate the heat can be equipped only in the service module (SVM) and 
it is not easy to transfer heat from the spinning PLM through the slip-ring to a non-spinning SVM.
The slip-ring introduces a single point whose failure would be critical for the mission.
Furthermore, a slip-ring might produce micro-vibration and could increase the detector noise
significantly.
For these reasons, we decided not to adopt the slip-ring and to rotate the whole spacecraft.

The spacecraft has a thrust tube at its center, which transfers the PLM launch load to the rocket. 
We will install the fuel tank inside the thrust tube to utilize the inner space effectively.
The insides of the side panels are used to mount various electric components of both the SVM and PLM\@.
PLM components are preferentially placed on the upper parts of the side panels, whereas SVM components are on the lower parts of the side panels.
The outer sides of the upper parts of the side panels are used to mount radiators, which radiate the heat dissipation
of the PLM, such as from the mechanical coolers and electronics boxes.

\begin{figure}[htb!]
\begin{center}
\includegraphics[width=1.0\linewidth]{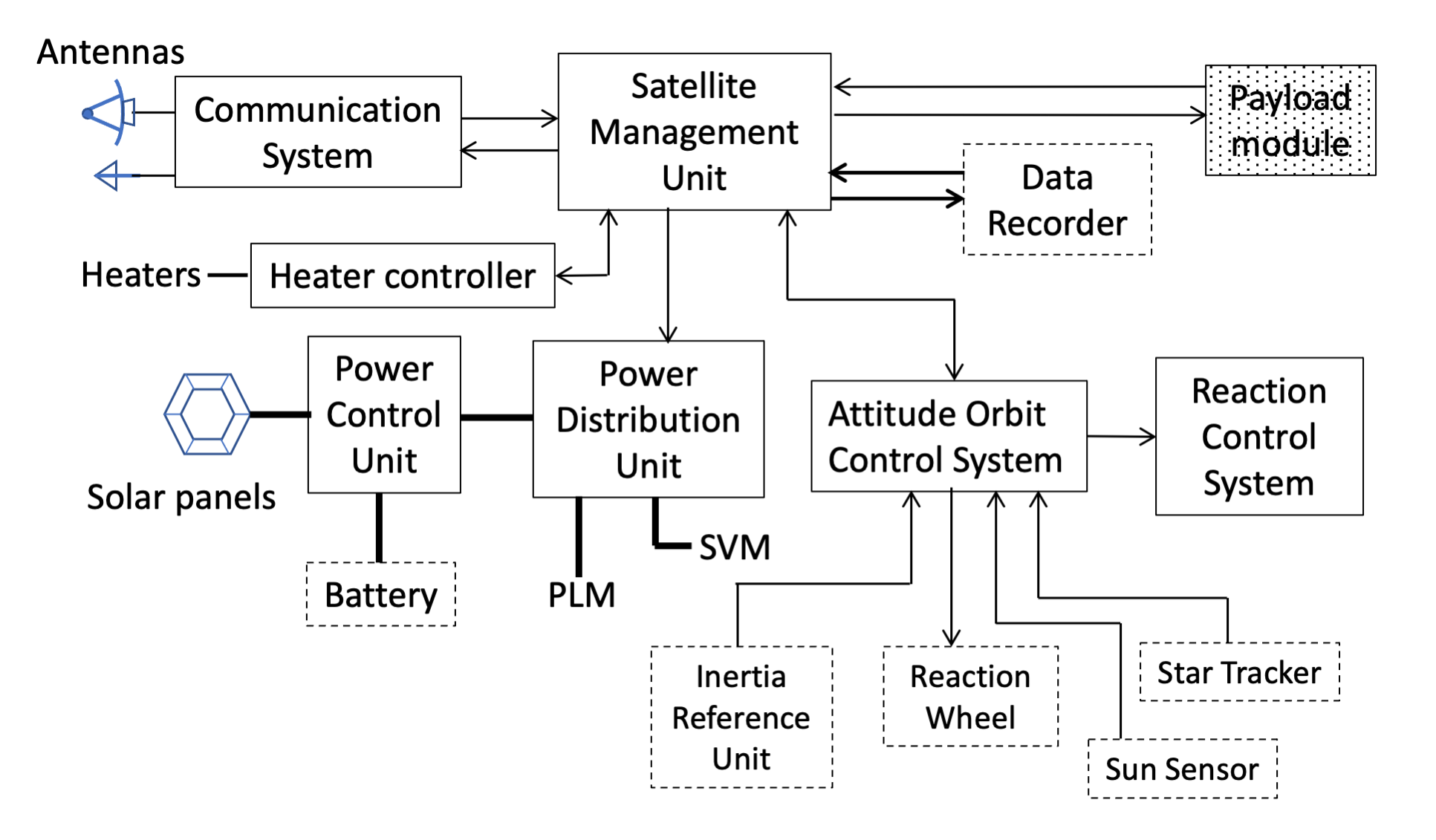}
\caption{Block diagram of the spacecraft for LiteBIRD. A box with broken lines represents electric equipment, while those with solid lines are subsystems composed of multiple equipment types.  Lines and arrows connecting boxes are only representative.}
\label{fig_spacecraftBlockDaigram}
\end{center}
\end{figure}
We show the block diagram of the spacecraft in Fig.~\ref{fig_spacecraftBlockDaigram}.
The LiteBIRD spacecraft takes a typical satellite configuration. 
Although the spacecraft spins, its attitude control system works like a 3-axis stabilized satellite to satisfy the attitude control and determination accuracy requirements. 
The low spin rate (nominal 0.05\,rpm, contingency 0.3\,rpm) makes this possible.

The spacecraft will have a total weight of 2.6\,t, including the fuel of approximately 400\,kg and with a total height of 5.3\,m.
Thus the current weight has a large margin compared to the rocket's capability.
We estimate the total power of the spacecraft to be 3.0\,kW\@.
The downlink rate will be $10\,{\rm Mbits}\,{\rm s}^{-1}$ in X-band and will transfer a total of 17.9\,GB of scientific data every day.
All these parameters are subject to change as the conceptual design of the satellite continues to be developed.

\section{Payload Module}
\label{sec:payload}
\begin{figure}[htb!]
\begin{center}
\includegraphics[width=1.0\linewidth]{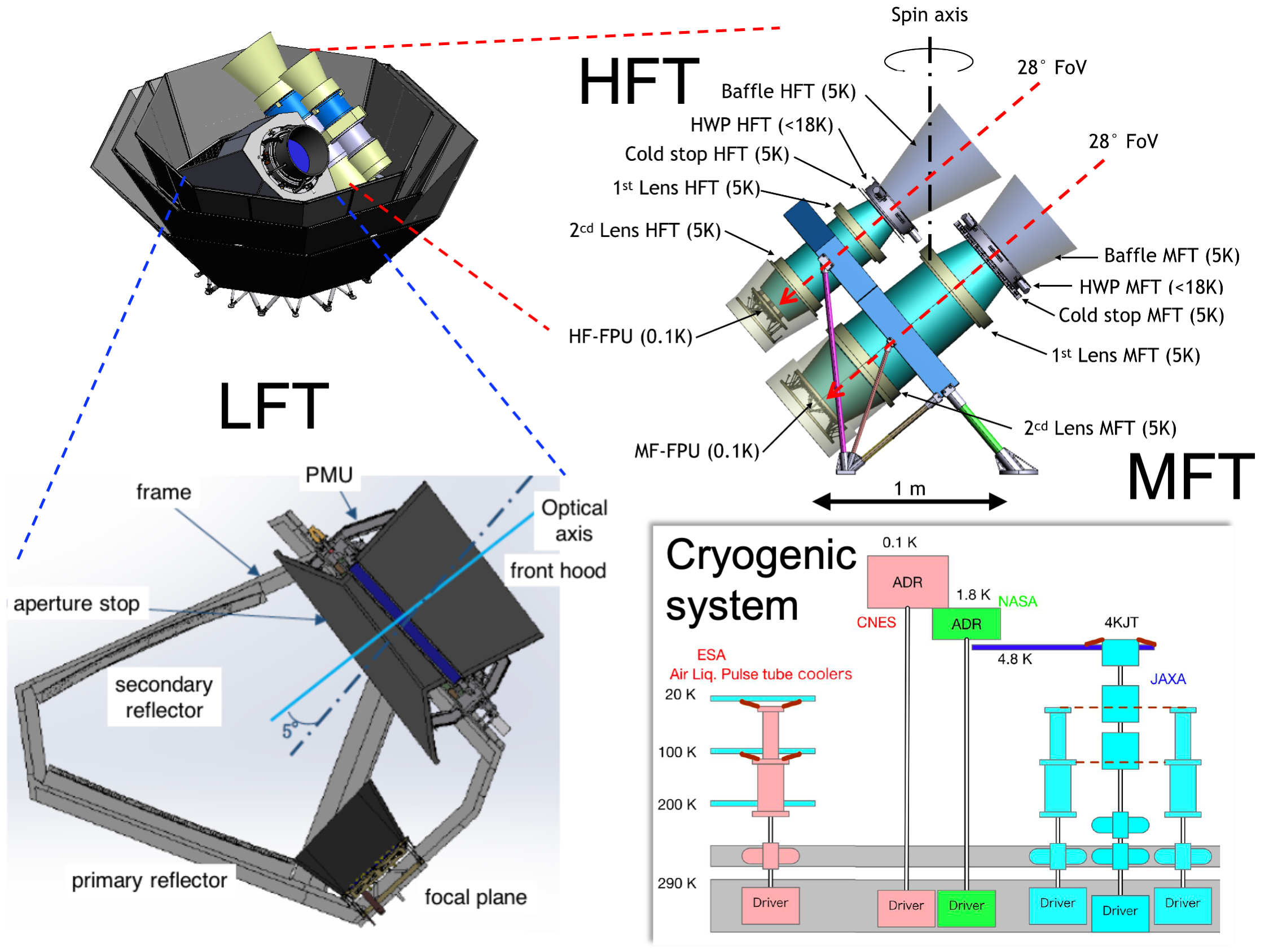}
\caption{Overview of the payload module (PLM).}
\label{fig_plm}
\end{center}
\end{figure}
%
\begin{figure}[htb!]
\centering
\includegraphics[width = 0.6\textwidth]{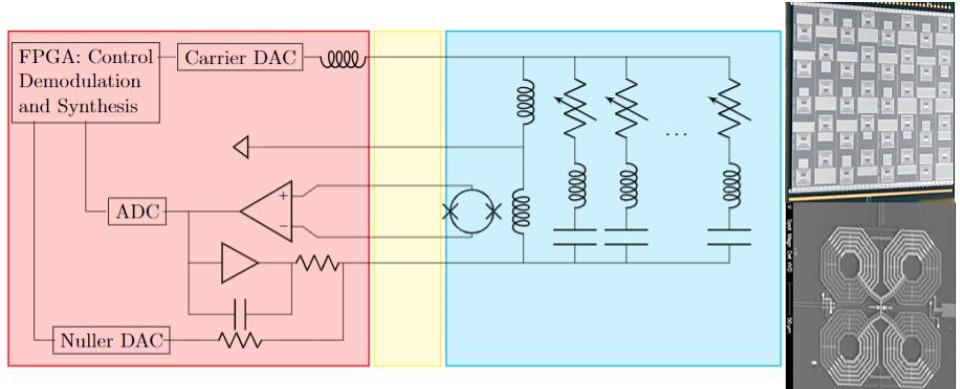}
\caption{
Schematic overview of the cryogenic readout system, with digital frequency-domain multiplexing (left) 
and images of a chip of 40 inductor-capacitor resonators (right-top) and a single gradiometric SQUID (right bottom).
The red section indicates the part of the circuit located at 300\,K, blue section is the part located on the 100-mK stage,
and yellow id the twisted pair wiring harness that connects them.}
\label{fig:dfmux}
\end{figure}

\noindent
Figure~\ref{fig_plm} shows an overview of the baseline-mission payload design of LiteBIRD.
The LiteBIRD payload module (PLM) consists of three telescopes -- at low, medium, and high frequencies -- with their respective focal planes and cryostructure cooled down to 0.1\,K. 
It also includes the global cooling chain from 300\,K to 5\,K and room-temperature elements, such as drivers and warm readout electronics of the detectors. 
We derive PLM requirements from the top-level requirement of achieving a tensor-to-scalar ratio error of $\delta r < 0.001$. 
This implies technical challenges for the PLM regarding sensitivity, optical properties, stability, or even compactness over a wide range of frequencies, from 34 to 448\,GHz.\footnote{The lowest (highest) frequency band has its center frequency at 40\,GHz (402\,GHz) and a fractional bandwidth of 30\,\% (23\,\%), giving the lower (upper) band edge at 34\,GHz (448\,GHz).}  On the other hand, the angular resolution requirement is not stringent, since we only need to cover the multipoles $2 \le \ell \le 200$. As a result, we obtain relaxed constraints on the telescopes' angular resolution (less than 80\,arcmin), but a robust control of the systematics is needed to minimize the $1/f$ noise. 

In this context, a critical technical choice made for LiteBIRD was to use as the first optical element a continuously-rotating half-wave plate (HWP) in the polarization modulation unit (PMU) for each telescope. 
The HWP allows us to distinguish between the instrumental polarization signal and the sky signal because the HWP modulates the latter signal only at a frequency of $4f_{\rm HWP}$. 
If we do not use the HWP, we need to take the signal difference between pairs of detectors that are mutually orthogonal in the polarization orientation.
This method is known to cause leakage from temperature to polarization if there are any differences in the beam, gain, or band-pass features between the two detectors.
We can significantly reduce the intensity-to-polarization leakage when we use the HWP, enabling us to measure the polarization using a single detector. 
The presence of the continuously-rotating HWP additionally performs an effective suppression of the $1/f$ noise. 
We carried out detailed trade-off studies between two cases, with and without the HWP, simulating  the polarization effects caused by the imperfection of the HWP to make a fair comparison. We found that the performance without the HWP is lower than in the case with the HWP, preventing us from satisfying the scientific requirement on $\delta r$. 
Hence, to guarantee appropriate thermal performances in terms of stability and minimal heat load, the three telescopes will be equipped with PMUs.
The revolution rate of each HWP is a function of the scan speed and the beam size.
We chose 46, 39, and 61 rpm for LFT, MFT, and HFT, respectively.

We have optimized the number of bands and their distributions over a wide range of frequency, from 34 to 448\,GHz, to deal with the following constraints.
\begin{enumerate}
\item The spectral bandwidth has to ensure the appropriate characterization of the expected complexity of the spectral energy distribution of the synchrotron and dust Galactic foregrounds, leading to 15 partially overlapping broad bands.
\item The limited frequency range of HWP materials (sapphire and metal-mesh) led us to split the entire spectral range into three telescopes.
\item The spectral mapping of the CO lines has to be optimized by rejecting such molecular lines from some of the bands and including them in others (notice that we decided not to use notch-filters, since we have demonstrated that the rotating HWP highly mitigates temperature-to-polarization leakage from CO-lines).
\item An overlap between bands and instruments was foreseen to mitigate systematic effects. 
\end{enumerate}
We ended up with the following distribution: a reflective telescope at low frequency, the LFT (34--161\,GHz); and two refractive telescopes at middle and high frequencies, the MFT (89--225\,GHz) and HFT (166--448\,GHz). We plan to mount the MFT and HFT on the same structure. They point in a different direction than the LFT; however, they cover the same circle over the sky when spinning. 
More details on the LFT are found in Ref.~\citenum{Sekimoto2020}, and on the MFT and HFT in Ref.~\citenum{Montier2020}.

The three telescopes' focal planes with a large field of view ($18^\circ\times 9^\circ$ for LFT, $28^\circ$ for MFT and HFT) are populated with multichroic polarized transition-edge sensor (TES) detectors (one to three bands per pixel). 
The multichroic technology allows a very compact design with sufficient flexibility to optimize the sensitivity per band required to improve the component separation. We have been using two detector technologies: lenslet-coupled detectors for the low- and medium-frequency telescopes; and horn-coupled detectors for the high-frequency telescope, for a total of 4339 detectors cooled down to 100\,mK. More details on the focal plane design and detector fabrication are described in Ref.~\citenum{Westbrook2020}. 
The readout electronics\cite{Dobbs2008,Montgomery2020} (Fig.~\ref{fig:dfmux}) takes advantage of the frequency-multiplexing scheme to accommodate this large set of detectors without losing information and with minimal power dissipation on the focal planes. 

%
\begin{figure}[htb!]
\begin{center}
\includegraphics[width=0.95\linewidth]{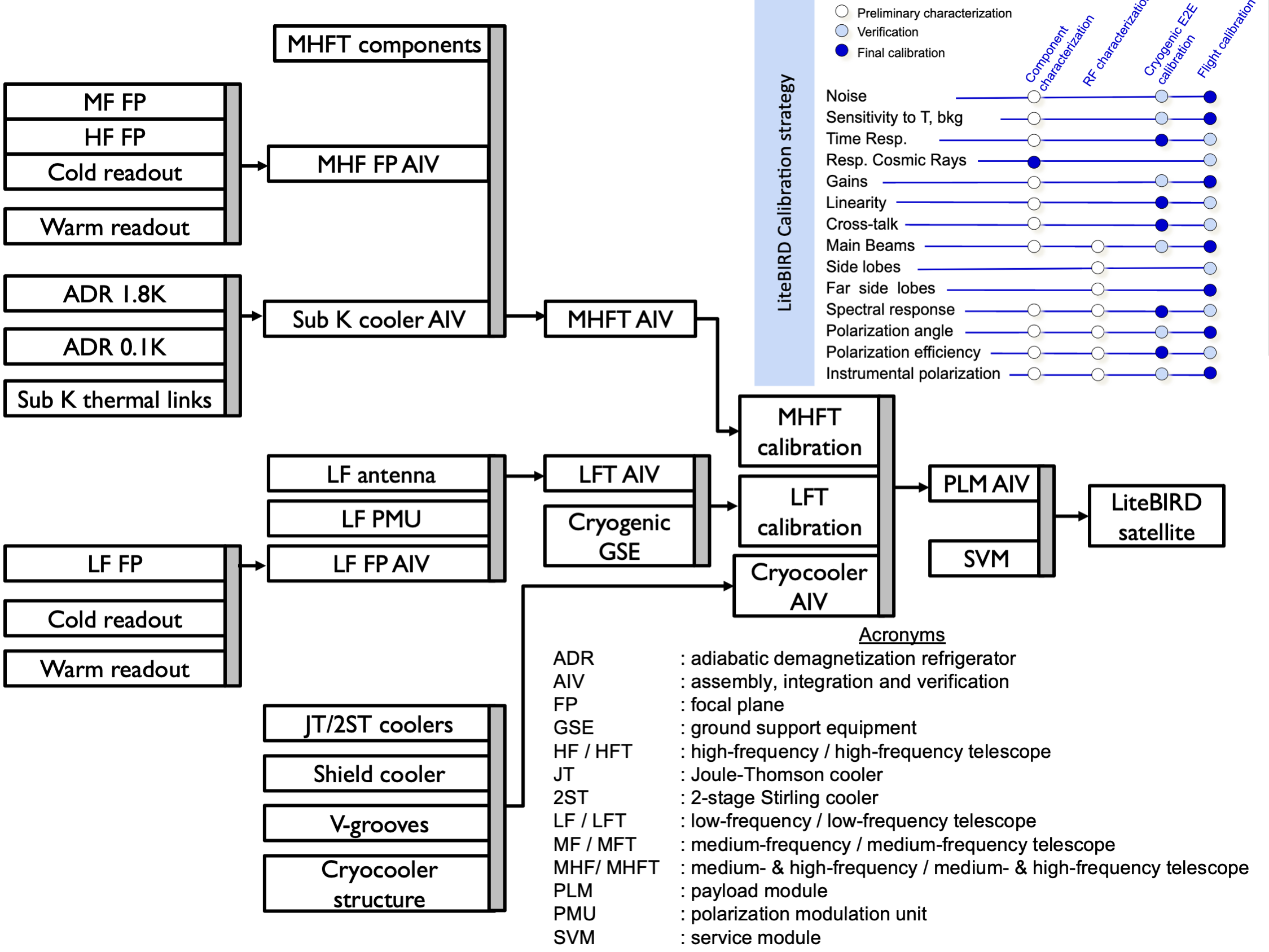}
\caption{Assembly, integration, verification (AIV), and pre-launch calibration of LiteBIRD. The inset in the top-right corner shows the current LiteBIRD calibration strategy.}
\label{fig:aivc}
\end{center}
\end{figure}

The instruments' temperature stability is another crucial point for CMB $B$-mode polarization probes for two reasons. Firstly, the temperature fluctuation of optical components contributes to noise stability and $1/f$ noise. Secondly, temperature variations of mechanical structures have a direct impact on pointing stability.
Hence the three LiteBIRD telescopes are thoroughly cooled to 4.8\,K to minimize the focal planes' heat load. The proposed 300-K to 4.8-K cryogenic chain for LiteBIRD adopts the architecture developed as part of the SPICA-SAFARI mission. This combines radiative cooling (V-grooves) down to 30\,K with mechanical cryocoolers to provide cooling to temperatures down to about 4.8\,K. 
In its current definition, a 15-K pulse tube cooler associated with three V-groove radiators, respectively at 160\,K, 90\,K, and 30\,K, intercept part of the thermal loads before a helium Joule-Thomson loop (4-K JT, 4He), pre-cooled by two two-stage Stirling coolers (100\,K and 20\,K).
Between the 4.8-K mechanical enclosure and the 0.1-K detectors,
all telescopes have intermediate cold stages at 1.8\,K and 0.3\,K. 
The 1.8-K cooler has three ADR stages operating in parallel, to provide a continuous cooling at 1.8\,K. We will use the controlled heat rejection of the 1.8-K cooler operation to damp the 4.8-K stage thermal oscillations. The sub-kelvin cooler consists of two ADR stages in parallel to provide stable and continuous cooling at 0.3\,K, combined with two other ADR stages in parallel at 0.1-K. We have optimized this cryochain design to ensure maximum stability of the focal planes' temperature and the optical elements of the telescopes. 

Figure~\ref{fig:aivc} shows the assembly, integration, verification (AIV), and pre-launch calibration of LiteBIRD.
We have chosen the integration scheme carefully to keep interfaces between different institutions and countries as simple as possible.
The inset on the top-right corner of Fig.~\ref{fig:aivc} shows our calibration strategy.
The plan is to derive a common approach for both instruments, LFT and MHFT, other than for
some specific exceptions. 
The requirements on the accuracy of the measurements of the instrumental parameters, which serve as inputs for the determination of the calibration strategy, are derived from detailed systematics studies (see Sect.~\ref{sec:outcomes}). 
The first step is to characterize the performance at the component level. 
These characterizations are part of the deliverables of the sub-systems. 
They will be based on the LiteBIRD specification and carried out before integration at the instrument level. 
We will also use the data from these characterizations to build an instrument model and forecast the in-flight performance as we develop the system.
Considering the beam calibration's specific challenges, we foresee a dedicated set of measurements, identified as `RF characterization' in the inset.
We will perform the instrument-level calibration for LFT and MHFT independently (in Japan and Europe, respectively) in a cold flight-like environment. 
We will perform a part of the final verification at the PLM level (system-level testing) when we assemble the LFT and MHFT with the satellite PLM and the SVM, together with the entire LiteBIRD cooling system.
As highlighted in the inset, we plan to rely on ground-calibration operations mostly for some parameters to ensure that they are accurate enough determined (to limit the impact on the systematic error on $r$); the spectral response is an example.
For other parameters (such as the main beam), the planned accuracy with flight data
should allow us to rely on flight data themselves. In such a case, we will use ground measurements as the first guess for preliminary analysis and, later, as a reference for verifications. 
It is worth noting that we do not exclude the option of solving for some systematic parameters as part of the map-making or component-separation processes. 
However, the calibration design philosophy does not rely on those post-analysis mitigations. Our strategy is to pursue the hardware developments of ground-segment equipment as the current baseline. 
In parallel, we explore the possibility of mitigating the systematics through post-analysis steps, which we currently view as a potential safeguard. 
We will refine the calibration plans, the error budget allocation for hardware development, and the post-flight analysis mitigation strategies as the project evolves.

\section{Operations concept}
\label{sec:operation}
\begin{figure}[htb!]
\centering
\includegraphics[width = 0.8\textwidth]{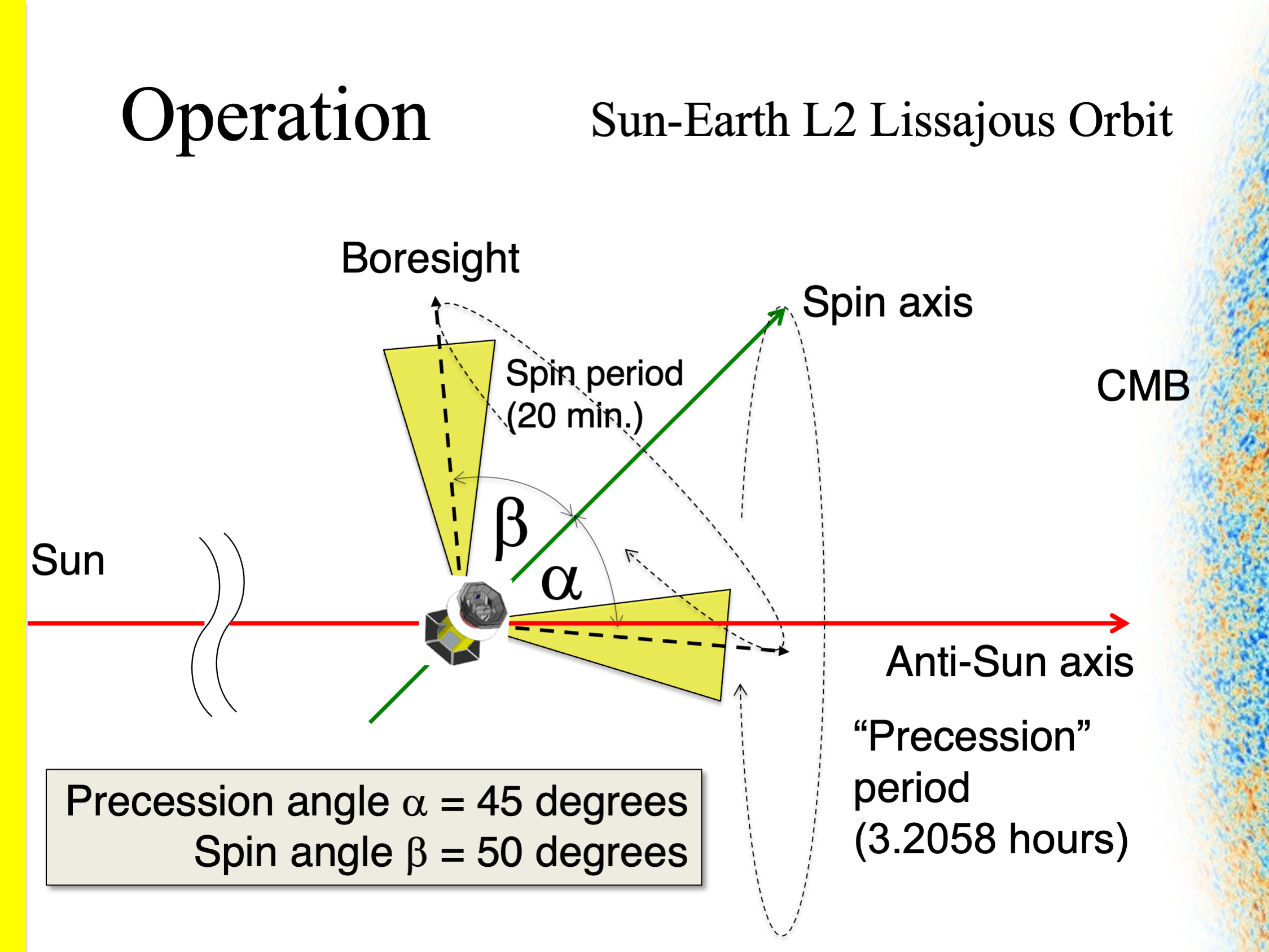}
\caption{Scan strategy of LiteBIRD in a Lissajous orbit around L2.}
\label{fig:scan_strategy}
\end{figure}
%
\begin{figure}[htb!]
\begin{center}
\includegraphics[width=0.7\linewidth]{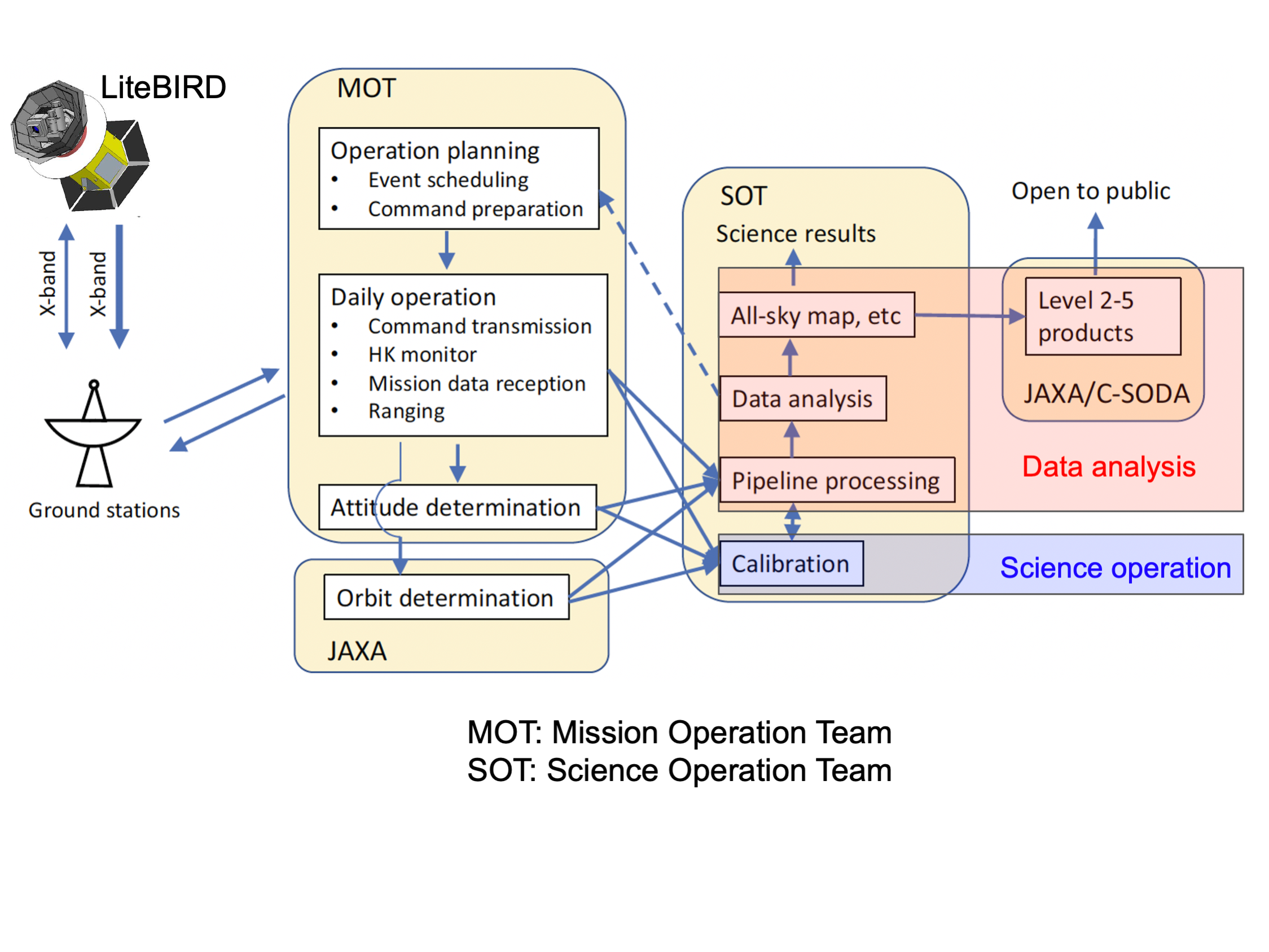}
\vspace{-1.0truecm}
\caption{Expected organization structure of the mission and science operations of LiteBIRD.}
\label{fig:sgs}
\end{center}
\end{figure}
We will use JAXA's H3 rocket to launch LiteBIRD and insert it into an orbit around the Sun-Earth Lagrangian point known as L2, to carry out full-sky surveys for three years.
A Lissajous orbit is our choice due mainly to the better thermal conditions compared to halo orbits.
In our scan strategy (Fig.~\ref{fig:scan_strategy}), the spacecraft spins about the spin axis at 0.05\,rpm, i.e., in 20\,minutes.
The spin axis itself also rotates or precesses around the anti-Sun axis;
however, the precession rate is much lower, with the current studies using a precession time of 3.2058 hours.
Here an irrational number is used to avoid systematics due to synchronization of the spin and precession.
The spin axis is canted $\alpha =$ 45$^\circ$ off the Sun-L2 axis, while 
the angle $\beta$ between the boresight and the spin axis is 50$^\circ$.

The requirements on the scan strategy include high thermal stability, 
good uniformity on the moving direction of boresight pointing
across each sky pixel (`attack angle' uniformity), 
near observational uniformity on sky pixels, broad daily sky coverage,
and short revisit times for each sky pixel.
These are all important to mitigate instrumental systematic uncertainties.

Figure~\ref{fig:sgs} shows the expected organizational structure of the mission and science operations of LiteBIRD.
JAXA is preparing a new GRound
station for deep-space Exploration And Telecommunication named GREAT\cite{JAXA_GREAT}, which should be operational before the launch of LiteBIRD.
The X-band transmission capability of GREAT will be sufficient to downlink all the data every day.
JAXA is also responsible for the management of the mission operation team, while
the science operations team (the SOT) is international and is responsible for analyzing observational data and producing scientific results. The SOT also works on the integration of the observational data and ground-calibration data.

\section{Expected scientific outcomes}
\label{sec:outcomes}
For cosmological forecasts, we have used the focal-plane parameters summarized in
Table~\ref{tbl:focalplane}.
%
\begin{table}[htb!]
\setlength\tabcolsep{4pt} 
\centering
\begin{tabular}{|r|r|r|r|r|r|r|r|r|}
\hline
Telescope	& Band  	& Center    & Frequency		& $\theta_{\rm FWHM}$ 	& Detector   	& Total     	& NET$^T_{\rm array}$ 	& $\omega^{-1/2}_P$ \\
          	& ID       	& Frequency & band [GHz]    & [arcmin]            	& pixel size 	& Number of 	& [$\mu$K$\sqrt{\rm s}$]   	& [$\mu$K-arcmin] \\ 
          	&         	& [GHz]     & (fraction)    &                     	& [mm]       	& Detectors 	&                     	&                 \\     
\hline
LFT	      	& 1			& 40 		& 12 (0.30)	 	& 70.5					& 32			& 48			& 18.50				  	& 37.42 \\ 		
\hline
LFT	      	& 2			& 50 		& 15 (0.30)	 	& 58.5					& 32			& 24			& 16.54				  	& 33.46 \\ 	
\hline
LFT	      	& 3			& 60 		& 14 (0.23)	 	& 51.1					& 32			& 48			& 10.54				  	& 21.31 \\ 	
\hline
LFT	      	& 4			& 68 		& 16 (0.23)	 	& (41.6, 47.1)			& (16, 32)		& (144, 24)		& (9.84, 15.70)		  	& (19.91, 31.77) \\
combined	&			&			&				&						&				&				& 8.34					& 16.87\\ 			
\hline
LFT	      	& 5			& 78 		& 18 (0.23)	 	& (36.9, 43.8)			& (16, 32)		& (144, 48)		& (7.69, 9.46)			& (15.55, 19.13) \\
combined	&			&			&				&						&				&				& 5.97					& 12.07\\ 		
\hline
LFT	      	& 6			& 89 		& 20 (0.23)	 	& (33.0, 41.5)			& (16, 32)		& (144, 24)		& (6.07, 14.22)			& (12.28, 28.77) \\
combined	&			&			&				&						&				&				& 5.58					& 11.30 \\
\hline
LFT/  		& 7			& 100 		& 23 (0.23)		& 30.2/					& 16/			& 144/			& 5.11/					& 10.34 \\
MFT			&			&			& 				& 37.8 					& 11.6 			& 366 			& 4.19 					& 8.48 \\
combined	&			&			&				&						&				&				& 3.24					& 6.56 \\			
\hline
LFT/		& 8			& 119 		& 36 (0.30)		& 26.3/					& 16/			& 144/			& 3.8/					& 7.69 \\
MFT			&			&			&				& 33.6					& 11.6			& 488			& 2.82					& 5.70 \\
combined	&			&			&				&						&				&				& 2.26					& 4.58 \\
\hline
LFT/		& 9			& 140 		& 42 (0.30)		& 23.7/					& 16/			& 144/			& 3.58/					& 7.25 \\
MFT			&			&			&				& 30.8					& 11.6			& 366			& 3.16					& 6.38 \\
combined	&			&			&				&						&				&				& 2.37					& 4.79 \\		
\hline
MFT	      	& 10		& 166		& 50 (0.30)		& 28.9					& 11.6			& 488			& 2.75					& 5.57 \\
\hline
MFT/   		& 11		& 195		& 59 (0.30)		& 28.0/					& 11.6/			& 366/			& 3.48/ 				& 7.05 \\
HFT			&			&			&				& 28.6					& 6.6			& 254			& 5.19					& 10.50 \\	
combined	&			&			&				&						&				&				& 2.89					& 5.85 \\
\hline
HFT	      	& 12		& 235		& 71 (0.30)		& 24.7					& 6.6			& 254			& 5.34					& 10.79 \\
\hline
HFT	      	& 13		& 280		& 84 (0.30)		& 22.5					& 6.6			& 254			& 6.82					& 13.80 \\
\hline
HFT	      	& 14		& 337		& 101 (0.30)	& 20.9					& 6.6			& 254			& 10.85					& 21.95 \\
\hline
HFT	      	& 15		& 402		& 92 (0.23)		& 17.9					& 5.7			& 338			& 23.45					& 47.45 \\
\hline
Total   	&			&			&				&						&				& 4508			&						& 2.16 \\
\hline
\end{tabular}
\caption{Focal-plane parameters of the 2020 baseline design of LiteBIRD. In the calculation, the aperture stop
temperature, mirrors for LFT, and lenses for HFT are at 5\,K.
The NET values include a margin (13\,\%), and the expected noise on the polarization signal on a sky pixel ($\omega^{-1/2}_P$) takes into account the end-to-end detector/readout yield of 80\,\%, as well as inefficiencies due to cosmic-ray hits (15\,\%) and ADR recycling (15\,\%).}
\label{tbl:focalplane}
\end{table}
We use the method described in Ref.~\citenum{Errard:2018ctl} for foreground cleaning.
One of the critical points is that we take spatial variations into account as much as possible. 
We separate the entire sky into 768 regions. In each sky region, we model the synchrotron radiation with a spectral index and its running, and dust emission with a spectral index and a modified blackbody temperature parameter. We treat Stokes $Q$ and $U$ polarization independently. As a result, we have eight parameters in each sky region and $8\times768 = 6144$ fit parameters in total.
%
\begin{figure}[htb!]
\centering
\includegraphics[width = 0.7\textwidth]{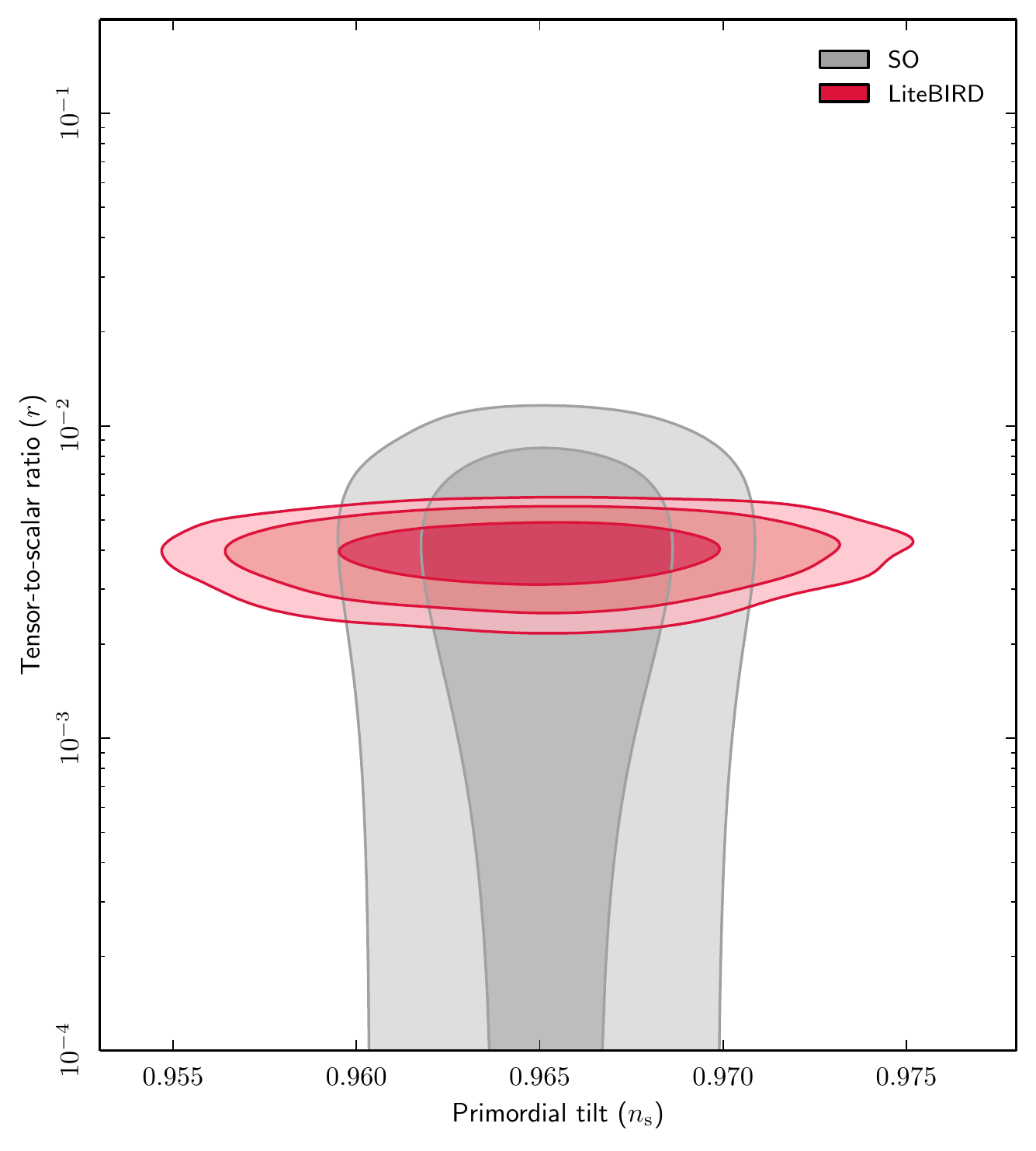}
\caption{Sensitivity contours of LiteBIRD and the Simons Observatory. Systematic uncertainties are not included here. The plot assumes that the actual value of $r$ is 0.004.
}
\label{fig:r_ns_contour}
\end{figure}
Our simulations yield $\sigma_{\rm stat} = 0.6\times 10^{-3}$ for $r=0$ as an input after foreground cleaning, with negligibly small bias. We also confirmed a consistent result from an alternative foreground cleaning method described in Ref.~\citenum{Andersen:2020jfq}.
Figure~\ref{fig:r_ns_contour} shows sensitivity contours of LiteBIRD.
As a comparison, we also show the expectation from a leading ground-based project in this decade, namely Simons Observatory\cite{Ade:2018sbj}.
Here one of the most promising inflationary models is assumed, specifically the Starobinsky model that predicts $r \simeq 0.004$.
We see the excellent discovery potential of LiteBIRD.
More details will be given in a comprehensive overview of LiteBIRD\cite{LiteBIRD_PTEP} that is in preparation.

We adopt a methodical approach for estimating systematic uncertainties.
We start by listing sources of systematic uncertainties, where we have
identified 70 items in 14 categories.
For each item, we allocate 1\,\% of the total budget.
This is based on our measurement requirement on the systematic error budget allocation (Lv2.07),
which states the following: 
`We shall decouple studies of each systematic error on $r$ as much as possible. Each component of systematic error on $r$ shall be less than 1\,\% of the total budget (Lv2.02), i.e., $\sigma_{\rm syst}$ from each component be less than $0.57 \times 10^{-5}$. In case an outstanding component is identified, however, it is allowed to allocate a particular budget for that item. If this happens, a careful investigation shall be done and a collaboration-wide agreement shall be made.'
We define our method for each source of systematic uncertainty, where some assumptions on the calibration methods
are also introduced.
As an example of such studies, we considered a source of systematic errors from half-wave plate imperfections. 
We model these imperfection based on a Mueller matrix from simulations of the rigorous coupled-wave analysis (RCWA).
We obtain leakage maps and resulting $B$-mode power from that and find that the result satisfies our requirements.
Another example is the systematic uncertainty due to cosmic-ray hits, which is described elsewhere.\cite{Tominaga2020,Stever2020}
%
In case we find some outstanding items, we will allocate particular budgets.
To obtain the total systematic error on $r$, $\sigma_{\rm syst}$, we make a sum in the map basis.
We iterate this procedure by adjusting error budget allocations until we achieve the goal. 
Finally, combining $\sigma_{\rm syst}$ with $\sigma_{\rm stat}$, we have confirmed that we meet the requirement of $\delta r < 0.001$ under some assumptions on the accuracy of our calibration methods\cite{LiteBIRD_PTEP}.

Here we do not assume any combination of LiteBIRD data with future astronomical observations that will likely be available when we analyze LiteBIRD data. We have reasonable expectations on improved CMB polarization data
in high $\ell$ regions from ground-based CMB projects and infrared survey data from space observations
for delensing, as well as low-frequency millimeter-wave observations on the ground for synchrotron radiation cleaning.\cite{LiteBIRD_PTEP} 
If we combine all these non-LiteBIRD data, we expect to obtain an even better constraint on the tensor-to-scalar ratio. We define this possibility as the `extra success' to enhance our science case beyond the `full success' of achieving $\delta r < 0.001$ for $2 \le \ell \le 200$.

We derive the system requirements of LiteBIRD from those for the tensor-to-scalar measurements alone.
Once we carry out the observations successfully, we can use LiteBIRD data to study many additional topics in cosmology, particle physics, and astronomy.
Examples include:
(1) characterization of $B$ modes and searches for source fields, such as scale-invariance, non-Gaussianity, and parity violation;
(2) power-spectrum features in polarization;
(3) cosmic-variance-limited measurements of large-scale (low-$\ell$) $E$ modes, with implications for the reionization history and the sum of neutrino masses;
(4) searches for cosmic birefringence;
(5) thermal Sunyaev-Zeldovich effects and relativistic corrections;
(6) elucidating large-angle anomalies; and
(7) Galactic astrophysics.
LiteBIRD has very targeted mission requirements, and at the same time will provide rich scientific outcomes.

\section{Summary}
\label{sec:summary}
LiteBIRD is a space mission for primordial cosmology and fundamental physics.
JAXA selected LiteBIRD in May 2019 as a strategic large-class (L-class) mission, with its expected launch in the 2020s using JAXA's H3 rocket.
The LiteBIRD Joint Study Group has more than 250 researchers from Japan, North America, and Europe.
LiteBIRD plans to map the CMB polarization over the full sky with unprecedented precision.
Its main scientific objective is to carry out a definitive search for the signal from cosmic inflation, either making a discovery or ruling out well-motivated inflationary models.
The measurements of LiteBIRD will also provide us with an insight into the quantum nature of gravity and other new physics beyond the standard models of particle physics and cosmology.
To satisfy its essential science requirement, $\delta r\,{<}\,0.001$ for $2\,{\leq}\,\ell\,{\leq}\,200$, LiteBIRD will perform full-sky surveys for three years at the Sun-Earth Lagrangian point L2. LiteBIRD will use 15 frequency bands between 34 and 448 GHz with three telescopes to achieve a total sensitivity of $2.16\,\mu$K-arcmin with a typical angular resolution of 0.5$^\circ$ at 100 GHz. 
We have completed the pre-phase-A2 concept development studies of LiteBIRD successfully.
Table~\ref{tbl:specifications} shows baseline specifications of LiteBIRD as the result of these studies.
\begin{table}[htb!]
\centering
\begin{tabular}{|p{10em}|p{35em}|} 
\hline
\bf Item & \bf Specification\\
\hline \hline
Science requirement & $\delta r < 0.001$ for $2 \leq \ell \leq 200$\\
\hline
Target launch year & 2029\\
\hline
Launch vehicle & JAXA H3\\
\hline
Observation type & All-sky CMB surveys\\
\hline
Observation time & 3 years\\
\hline
Orbit & L2 Lissajous orbit\\
\hline
Scan and  		& $\cdot$ Spin and precession (precession angle $\alpha = 45^\circ$, spin angle $\beta = 50^\circ$)\\
data recording	& $\cdot$ Spin period = 20\,minutes, precession period = 3.2058\,hours\\
				& $\cdot$ PMU revolution rate = 46/39/61\,rpm for LFT/MFT/HFT\\
				& $\cdot$ Sampling rate = 19\,Hz\\
\hline
Observing frequencies  & 34--448\,GHz\\
\hline
Number of bands & 15 \\
\hline
Polarization sensitivity & $2.16\,\mu$K-arcmin (after 3 years) \\
\hline
Angular resolution & 0.5$^\circ$ at 100\,GHz (FWHM for LFT) \\
\hline
Mission instruments & $\cdot$ Superconducting detector arrays \\
                    & $\cdot$ Crossed-Dragone mirrors (LFT) + two refractive telescopes (MFT and HFT) \\
                    & $\cdot$ PMU with continously-rotating HWP on each telescope\\
                    & $\cdot$ 0.1-K cooling chain (ST/JT/ADR) \\
\hline
Data size & $17.9\,{\rm GB}\,{\rm day}^{-1}$ \\
\hline
Mass & 2.6\,t \\
\hline
Power & 3.0\,kW \\
\hline
\end{tabular}
\caption{Main specifications of LiteBIRD. Parameters are from the LiteBIRD pre-phase-A2 concept development studies
and additional studies in 2020 as preparation for the system-requirements review.}
\label{tbl:specifications}
\end{table}

\acknowledgments 
This work is supported in Japan by ISAS/JAXA for Pre-Phase A2 studies, by the acceleration program of JAXA research and development directorate, by the World Premier International Research Center Initiative (WPI) of MEXT, by the JSPS Core-to-Core Program of A. Advanced Research Networks, and by JSPS KAKENHI Grant Numbers JP15H05891, JP17H01115, and JP17H01125. The Italian LiteBIRD phase A contribution is supported by the Italian Space Agency (ASI Grants No. 2020-9-HH.0 and 2016-24-H.1-2018), the National Institute for Nuclear Physics (INFN) and the National Institute for Astrophysics (INAF). The French LiteBIRD phase A contribution is supported by the Centre National d’Etudes Spatiale (CNES), by the Centre National de la Recherche Scientifique (CNRS), and by the Commissariat à l’Energie Atomique (CEA). The Canadian contribution is supported by the Canadian Space Agency. The US contribution is supported by NASA grant no. 80NSSC18K0132. 
Norwegian participation in LiteBIRD is supported by the Research Council of Norway (Grant No. 263011). The Spanish LiteBIRD phase A contribution is supported by the Spanish Agencia Estatal de Investigación (AEI), project refs. PID2019-110610RB-C21 and AYA2017-84185-P. Funds that support the Swedish contributions come from the Swedish National Space Agency (SNSA/Rymdstyrelsen) and the Swedish Research Council (Reg. no. 2019-03959). The German participation in LiteBIRD is supported in part by the Excellence Cluster ORIGINS, which is funded by the Deutsche Forschungsgemeinschaft (DFG, German Research Foundation) under Germany’s Excellence Strategy (Grant No. EXC-2094 - 390783311). This research used resources of the Central Computing System owned and operated by the Computing Research Center at KEK, as well as resources of the National Energy Research Scientific Computing Center, a DOE Office of Science User Facility supported by the Office of Science of the U.S. Department of Energy.

\begin{center}
\end{center}
\bibliography{report_litebird_spie2020}

\begin{thebibliography}{10}

\bibitem{Hazumi:2008zz}
Hazumi, M., ``{Jumping into CMB polarization measurements: A new group at
  KEK},'' {\em AIP Conf. Proc.}~{\bf 1040}(1),  78--88 (2008).

\bibitem{Hazumi:2011zz}
Hazumi, M., ``{Future CMB polarization measurements and Japanese
  contributions},'' {\em Prog. Theor. Phys. Suppl.}~{\bf 190},  75--89 (2011).

\bibitem{Hazumi:2012gjy}
Hazumi, M. et~al., ``{LiteBIRD: a small satellite for the study of B-mode
  polarization and inflation from cosmic background radiation detection},''
  {\em Proc. SPIE Int. Soc. Opt. Eng.}~{\bf 8442},  844219 (2012).

\bibitem{Matsumura:2013aja}
Matsumura, T. et~al., ``{Mission design of LiteBIRD},'' {\em J. Low Temp.
  Phys.}~{\bf 176},  733 (2014).

\bibitem{Matsumura:2014fte}
Matsumura, T. et~al., ``{LiteBIRD: mission overview and design tradeoffs},''
  {\em Proc. SPIE Int. Soc. Opt. Eng.}~{\bf 9143},  91431F (2014).

\bibitem{Kamionkowski:1996zd}
Kamionkowski, M., Kosowsky, A., and Stebbins, A., ``{A Probe of primordial
  gravity waves and vorticity},'' {\em Phys. Rev. Lett.}~{\bf 78},  2058--2061
  (1997).

\bibitem{Seljak:1996gy}
Seljak, U. and Zaldarriaga, M., ``{Signature of gravity waves in polarization
  of the microwave background},'' {\em Phys. Rev. Lett.}~{\bf 78},  2054--2057
  (1997).

\bibitem{Zaldarriaga:1996xe}
Zaldarriaga, M. and Seljak, U., ``{An all sky analysis of polarization in the
  microwave background},'' {\em Phys. Rev. D}~{\bf 55},  1830--1840 (1997).

\bibitem{Kamionkowski:1996ks}
Kamionkowski, M., Kosowsky, A., and Stebbins, A., ``{Statistics of cosmic
  microwave background polarization},'' {\em Phys. Rev. D}~{\bf 55},
  7368--7388 (1997).

\bibitem{Sekimoto2020}
Sekimoto, Y. and the LiteBIRD Joint Study~Group, ``{Wide field-of-view design
  of low frequency telescope on CMB B-mode polarization satellite LiteBIRD},''
  Space Telescopes and Instrumentation 2020: Optical, Infrared, and Millimeter
  Wave (2020 in preparation).

\bibitem{Montier2020}
Montier, L. and the LiteBIRD Joint Study~Group, ``{Overview of the Medium- and
  High-Frequency Telescopes of the LiteBIRD satellite mission},'' Space
  Telescopes and Instrumentation 2020: Optical, Infrared, and Millimeter Wave
  (2020 in preparation).

\bibitem{Westbrook2020}
Westbrook, B., Raum, C., and Suzuki, A., ``{Detector fabrication development
  for the LiteBIRD satellite mission},'' in [{\em Space Telescopes and
  Instrumentation 2020: Optical, Infrared, and Millimeter
  Wave}{\nolinebreak\hspace{0.1em}]},  (2020 in preparation).

\bibitem{Sakurai2020}
Sakurai, Y. et~al., ``{Breadboard model of polarization modulator unit based on
  a continuous rotating half-wave plate for low frequency telescope of LiteBIRD
  space mission},'' in [{\em Space Telescopes and Instrumentation
  2020}{\nolinebreak\hspace{0.1em}]},  SPIE (2020).

\bibitem{HTakakura2020}
Takakura, H., Sekimoto, Y., Inatani, J., Kashima, S., and Sugimoto, M.,
  ``Polarization angle measurement of litebird low frequency telescope scaled
  model,'' in [{\em Space Telescopes and Instrumentation
  2020}{\nolinebreak\hspace{0.1em}]},  SPIE (2020).

\bibitem{Tsuji2020}
Tsuji, M., Tsujimoto, M., Sekimoto, Y., Dotani, T., and Shiraishi, M.,
  ``{Simulating electromagnetic transfer function from the transmission
  antennae to the sensors vicinity in LiteBIRD},'' Space Telescopes and
  Instrumentation 2020: Optical, Infrared, and Millimeter Wave (2020 in
  preparation).

\bibitem{Tominaga2020}
Tominaga, M., Tsujimoto, M., Stever, S., Ghigna, T., Ishino, H., and Ebisawa,
  K., ``{Cosmic ray glitch predictions , physical modeling , and overall effect
  on the LiteBIRD space mission ( 2 )},'' in [{\em Space Telescopes and
  Instrumentation 2020: Optical, Infrared, and Millimeter
  Wave}{\nolinebreak\hspace{0.1em}]},  (2) (2020 in preparation).

\bibitem{lamagna2021}
{Lamagna}, L., J.E., G., {Imada}, H., {Hargrave}, P., {Franceschet}, C.,
  {De~Petris}, M., {Austermann}, J., {Bounissou}, S., {Columbro}, F.,
  {de~Bernardis}, P., {Henrot-Versillé}, S., {Hubmayr}, J., {Jaehnig}, G.,
  {Keskitalo}, R., {Maffei}, B., {Masi}, S., {Matsumura}, T., {Montier}, L.,
  {Mot}, B., {Noviello}, F., {O'Sullivan}, C., {Paiella}, A., {Pisano}, G.,
  {Realini}, S., {Ritacco}, A., {Savini}, G., {Suzuki}, A., {Trappe}, N., and
  {Winter}, B., ``{The optical design of the LiteBIRD Middle and High Frequency
  Telescope},'' in [{\em Space Telescopes and Instrumentation 2020: Optical,
  Infrared, and Millimeter Waves}{\nolinebreak\hspace{0.1em}]},   {\bf
  11443-283}, Int. Soc. for Optics and Photonics, SPIE (2021).

\bibitem{Hinshaw:2012aka}
Hinshaw, G. et~al., ``{Nine-Year Wilkinson Microwave Anisotropy Probe (WMAP)
  Observations: Cosmological Parameter Results},'' {\em Astrophys. J.
  Suppl.}~{\bf 208},  19 (2013).

\bibitem{Bennett:2012zja}
Bennett, C. et~al., ``{Nine-Year Wilkinson Microwave Anisotropy Probe (WMAP)
  Observations: Final Maps and Results},'' {\em Astrophys. J. Suppl.}~{\bf
  208},  20 (2013).

\bibitem{Ade:2017uvt}
Ade, P. et~al., ``{A Measurement of the Cosmic Microwave Background $B$-Mode
  Polarization Power Spectrum at Sub-Degree Scales from 2 years of POLARBEAR
  Data},'' {\em Astrophys. J.}~{\bf 848}(2),  121 (2017).

\bibitem{Henning:2017nuy}
Henning, J. et~al., ``{Measurements of the Temperature and E-Mode Polarization
  of the CMB from 500 Square Degrees of SPTpol Data},'' {\em Astrophys.
  J.}~{\bf 852}(2),  97 (2018).

\bibitem{Ade:2018gkx}
Ade, P. et~al., ``{BICEP2 / Keck Array x: Constraints on Primordial
  Gravitational Waves using Planck, WMAP, and New BICEP2/Keck Observations
  through the 2015 Season},'' {\em Phys. Rev. Lett.}~{\bf 121},  221301 (2018).

\bibitem{Aghanim:2019ame}
Aghanim, N. et~al., ``{Planck 2018 results. V. CMB power spectra and
  likelihoods},'' {\em Astron. Astrophys.}~{\bf 641},  A5 (2020).

\bibitem{Sayre:2019dic}
Sayre, J. et~al., ``{Measurements of B-mode Polarization of the Cosmic
  Microwave Background from 500 Square Degrees of SPTpol Data},'' {\em Phys.
  Rev. D}~{\bf 101}(12),  122003 (2020).

\bibitem{Aiola:2020azj}
Aiola, S. et~al., ``{The Atacama Cosmology Telescope: DR4 Maps and Cosmological
  Parameters},'' {\em arXiv e-prints} ,  arXiv:2007.07288 (July 2020).

\bibitem{Choi:2020ccd}
Choi, S.~K. et~al., ``{The Atacama Cosmology Telescope: A Measurement of the
  Cosmic Microwave Background Power Spectra at 98 and 150 GHz},'' {\em arXiv
  e-prints} ,  arXiv:2007.07289 (July 2020).

\bibitem{Adachi:2020knf}
Adachi, S. et~al., ``{A measurement of the CMB E-mode angular power spectrum at
  subdegree scales from 670 square degrees of POLARBEAR data},'' {\em
  Astrophys. J.}~{\bf 904}(1),  65 (2020).

\bibitem{Tristram2020}
{Tristram}, M., {Banday}, A.~J., {G{\'o}rski}, K.~M., {Keskitalo}, R.,
  {Lawrence}, C.~R., {Andersen}, K.~J., {Barreiro}, R.~B., {Borrill}, J.,
  {Eriksen}, H.~K., {Fernandez-Cobos}, R., {Kisner}, T.~S.,
  {Mart{\'\i}nez-Gonz{\'a}lez}, E., {Partridge}, B., {Scott}, D., {Svalheim},
  T.~L., {Thommesen}, H., and {Wehus}, I.~K., ``{Planck constraints on the
  tensor-to-scalar ratio},'' {\em arXiv e-prints} ,  arXiv:2010.01139 (Oct.
  2020).

\bibitem{Kamionkowski:2015yta}
Kamionkowski, M. and Kovetz, E.~D., ``{The Quest for B Modes from Inflationary
  Gravitational Waves},'' {\em Ann. Rev. Astron. Astrophys.}~{\bf 54},
  227--269 (2016).

\bibitem{Linde:2016hbb}
Linde, A., ``{Gravitational waves and large field inflation},'' {\em J. Cosm.
  Astropart. Phys.}~{\bf 02},  006 (2017).

\bibitem{Diego-Palazuelos2020}
{Diego-Palazuelos}, P., {Vielva}, P., {Mart{\'\i}nez-Gonz{\'a}lez}, E., and
  {Barreiro}, R.~B., ``{Comparison of delensing methodologies and assessment of
  the delensing capabilities of future experiments},'' {\em J. Cosm. Astropart.
  Phys.}~{\bf 2020},  058 (Nov. 2020).

\bibitem{JAXA_H3}
{Japan Aerospace Exploration Agency (JAXA)}, ``{H3 Launch Vehicle}.''
  \url{https://global.jaxa.jp/projects/rockets/h3/}.

\bibitem{Dobbs2008}
{Dobbs}, M., {Bissonnette}, E., and {Spieler}, H., ``{Digital Frequency Domain
  Multiplexer for Millimeter-Wavelength Telescopes},'' {\em IEEE Transactions
  on Nuclear Science}~{\bf 55},  21--26 (Jan. 2008).

\bibitem{Montgomery2020}
Montgomery, J., {\em {Digital Frequency Domain Multiplexing readout: design and
  performance of the SPT-3G instrument and LiteBIRD satellite readout}}, PhD
  thesis, McGill University (2020).

\bibitem{JAXA_GREAT}
{Japan Aerospace Exploration Agency (JAXA)}, ``{GREAT, Ground Station for Deep
  Space Exploration and Telecommunication}.''
  \url{https://global.jaxa.jp/projects/sas/great/}.

\bibitem{Errard:2018ctl}
Errard, J. and Stompor, R., ``{Characterizing bias on large scale CMB B-modes
  after galactic foregrounds cleaning},'' {\em Phys. Rev. D}~{\bf 99}(4),
  043529 (2019).

\bibitem{Andersen:2020jfq}
Andersen, K. et~al., ``{BeyondPlanck I. Global Bayesian analysis of the Planck
  Low Frequency Instrument data},'' in [{\em {BeyondPlanck Release
  Conference}}{\nolinebreak\hspace{0.1em}]},  (11 2020).

\bibitem{Ade:2018sbj}
Ade, P. et~al., ``{The Simons Observatory: Science goals and forecasts},'' {\em
  JCAP}~{\bf 02},  056 (2019).

\bibitem{LiteBIRD_PTEP}
{The LiteBIRD Joint Study Group}, ``{Probing Cosmic Inflation with the LiteBIRD
  Cosmic Microwave Background Polarization Survey},'' {\em Progress in
  Theoretical and Experimental Physics} ,  in preparation.

\bibitem{Stever2020}
Stever, S.~L., Ghigna, T., Tominaga, M., Tsujimoto, M., Minami, Y., Sugiyama,
  S., Kato, A., Matsumura, T., Ishino, H., and Hazumi, M., ``{Simulations of
  systematic effects arising from cosmic rays in the LiteBIRD space telescope,
  and effects on the measurements of CMB $B$ modes},'' J. Cosmol. Astrophys.
  (2020 in preparation).

\end{thebibliography}
\bibliographystyle{spiebib} 

\end{document}